\documentclass[aps,prc,groupedaddress,showpacs,manuscript]{revtex4}

\usepackage{graphicx}

\begin{document}

\title{Probing the density dependence of the symmetry potential at low and high densities}

\author {Qingfeng Li$\, ^{1}$\footnote{Fellow of the Alexander von Humboldt Foundation.}
\email[]{Qi.Li@fias.uni-frankfurt.de}, Zhuxia Li$\, ^{1,2}$
\email[]{lizwux@iris.ciae.ac.cn}, Sven Soff$\, ^{3}$, Marcus
Bleicher$\, ^{3}$, and Horst St\"{o}cker$\, ^{1,3}$}
\address{
1) Frankfurt Institute for Advanced Studies (FIAS), Johann Wolfgang Goethe-Universit\"{a}t, Max-von-Laue-Str.\ 1, D-60438 Frankfurt am Main, Germany\\
2) China Institute of Atomic Energy, P.O.\ Box 275 (18),
Beijing 102413, P.R.\ China\\
3) Institut f\"{u}r Theoretische Physik, Johann Wolfgang Goethe-Universit\"{a}t, Max-von-Laue-Str.\ 1, D-60438 Frankfurt am Main, Germany\\
 }


\begin{abstract}
We investigate the sensitivity of several observables to the
density dependence of the symmetry potential within the microscopic
transport model UrQMD (ultrarelativistic quantum molecular dynamics model). The same systems are used to probe the symmetry potential at
both low and high densities. The influence of the symmetry
potentials on the yields of $\pi^{-}$, $\pi^{+}$, the
$\pi^{-}/\pi^{+}$ ratio, the $n/p$ ratio of free nucleons and the
$t/^3$He ratio are studied for neutron-rich heavy ion collisions
($^{208}Pb+^{208}Pb$, $^{132}Sn+^{124}Sn$, $^{96}Zr+^{96}Zr$) at
$E_b=0.4A\,{\rm GeV}$. We find that these multiple probes provides comprehensive information on the
density dependence of the symmetry potential.
\end{abstract}


\pacs{24.10.Lx, 25.75.Dw, 25.75.-q} \maketitle

The nuclear equation of state (EoS) for
isospin-asymmetric nuclear systems has recently attracted a lot of
attention. The EoS for isospin-asymmetric nuclear matter can be
described approximately by the parabolic law (see, e.g.,
\cite{Bom91})
\begin{equation}
e(\rho,\delta)=e_{0}(\rho,0)+e_{\rm sym}(\rho)\delta^{2},
\end{equation}
where $\delta=(\rho_{n}-\rho_{p})/(\rho_{n}+\rho_{p})$ is the
isospin asymmetry defined by the neutron ($\rho_n$) and proton
($\rho_p$) densities. $e_{0}$ is the energy per nucleon for
symmetric nuclear matter and $e_{\rm sym}(\rho)$ is the bulk
symmetry energy. The nuclear symmetry energy term $e_{\rm
sym}(\rho)$ is very important for the understanding of many
interesting astrophysical phenomena, but it is also plagued by
large uncertainties. An extreme variation of the EoS is observed
for neutron-rich matter by using different versions of the Skyrme
interaction as well as other non-relativistic effective
interactions \cite{Br00}. The EoS of neutron-rich matter can be
studied using a density-dependent relativistic hadron mean-field theory
like the DDH3$\rho\delta$ and DDH$\rho$ models \cite{Gai04}. The
density dependence of the symmetry potential deduced from
relativistic and non-relativistic interactions might be different.
Therefore, a comparison of the isospin effect in isospin-asymmetric
heavy ion collisions (HICs) with both relativistic and non-relativistic
effective interactions is important. Furthermore, most of
the studies on sensitive probes of the density dependence of the
symmetry potential are concerned with the behavior in the low-density or in the
high-density region separately. Thus the simultaneous study of observables, which are sensitive to both the
low-density and the high-density part of symmetry potential,
in one reaction system is necessary. This allows to
extract the full information on the density dependence of the
symmetry potential which relates to specific effective
interactions.

To study the isospin effects in intermediate energy
HICs, heavy stable nuclei with an isospin
asymmetry as large as possible should be used. They have the advantage that both, the availability and the beam intensities are much
higher than for radioactive beams and, hence, high statistics
experimental data can be obtained. This is necessary to study
the relatively subtle effects of different symmetry potentials. The
reaction $^{208}Pb+^{208}Pb$, for example, has an
isospin-asymmetry $\delta\simeq 0.212$, i.e., a neutron to proton ratio
$N/Z\simeq 1.54$, which is already
quite large \cite{Xiao05}.

\begin{figure}
\includegraphics[angle=0,width=0.8\textwidth]{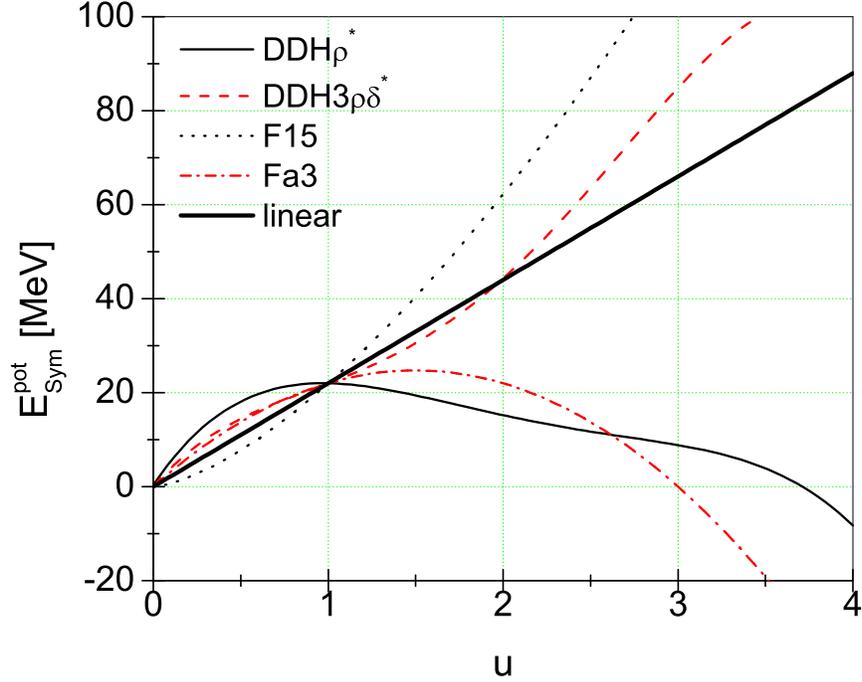}
\caption{Parametrizations of the nuclear symmetry potential
DDH$\rho^*$, DDH3$\rho\delta^*$, F15, Fa3, and the linear one as a
function of the reduced density $u$.} \label{fig1}
\end{figure}

In a previous work, we have adopted the following forms of the
symmetry potential \cite{Li05}
 \begin{equation}
F(u)=\left\{
\begin{array}{l}
F_1=u^\gamma \ , \hspace{1cm}  {\rm with} \hspace{0.5cm} \gamma>0 \\
F_2=u\cdot\frac{a-u}{a-1} \ , \hspace{1cm} {\rm with}
\hspace{0.5cm} a>1
\end{array}
\right. ,\label{fu}
\end{equation}
to mimic the strong variation of the density dependence of the
symmetry potential at high densities as given by the Skyrme-type
interactions. Here, $u=\rho/\rho_0$ is the reduced density and $a$
is the reduced critical density. When $u>a$, the symmetry
potential energy is negative in $F_2$. We choose $\gamma=1.5$ ($a=3$) for
$F_1$ ($F_2$) and name it F15 (Fa3). In the present work we further adopt
two other forms of the symmetry potential based on the
relativistic mean field theory, namely DDH3$\rho\delta^*$ and
DDH$\rho^*$. Both are based on the extended QHD model (see Ref.\
\cite{Gai04}). The density dependence of the symmetry potentials
for the DDH3$\rho\delta$ and DDH$\rho$ models is due to the
density dependence of the coupling constants (g$_{\sigma N}$,
g$_{\omega N}$, g$_{\rho N}$, and g$_{\delta N}$). Fig.\ \ref{fig1}
shows the density dependences of these symmetry potentials as well
as the linear case. For comparison, we fix the symmetry potential
energy at normal nuclear density for all four forms of the
symmetry potential. The symmetry energy coefficient is adopted to be $S_0=34$ MeV \cite{Vre03,Dal04}. It is seen from Fig. \ref{fig1}
that, for $u<1$, DDH3$\rho\delta^*$ is very close to Fa3, both of
them lie between DDH$\rho^*$ and F15. For $u>1$, they
show a rather different behavior and approach F15 and DDH$\rho^*$,
respectively.

In the present work, central collisions  are studied for the three
neutron-rich systems $^{208}Pb+^{208}Pb$ ($\delta \simeq 0.212$),
$^{132}Sn+^{124}Sn$ ($\delta \simeq 0.218$), and $^{96}Zr+^{96}Zr$
($\delta \simeq 0.167$) at $E_b=0.4A\,{\rm GeV}$. The ultrarelativistic quantum molecular dynamics model (UrQMD, version 1.3) \cite{Bass98,Bleicher99,Web03,Bra04} is adopted
for the calculations including the Coulomb potential for
charged pions. A 'hard' Skyrme-type EoS ($K=300$ MeV) without
momentum dependence is used for the calculations.  It should be
noted that the uncertainty of the isospin independent EoS is still
large (see Refs. \cite{Vre03,Riz04}), and the contribution of the momentum dependence in EoS is also important to the dynamics of HICs (see Ref. \cite{LiB04}), both of them  influence the results but do not change the effect of the density dependence of symmetry potential on the ratios of different charged
particles strongly \cite{LiB05}.

\begin{figure}
\includegraphics[angle=0,width=0.8\textwidth]{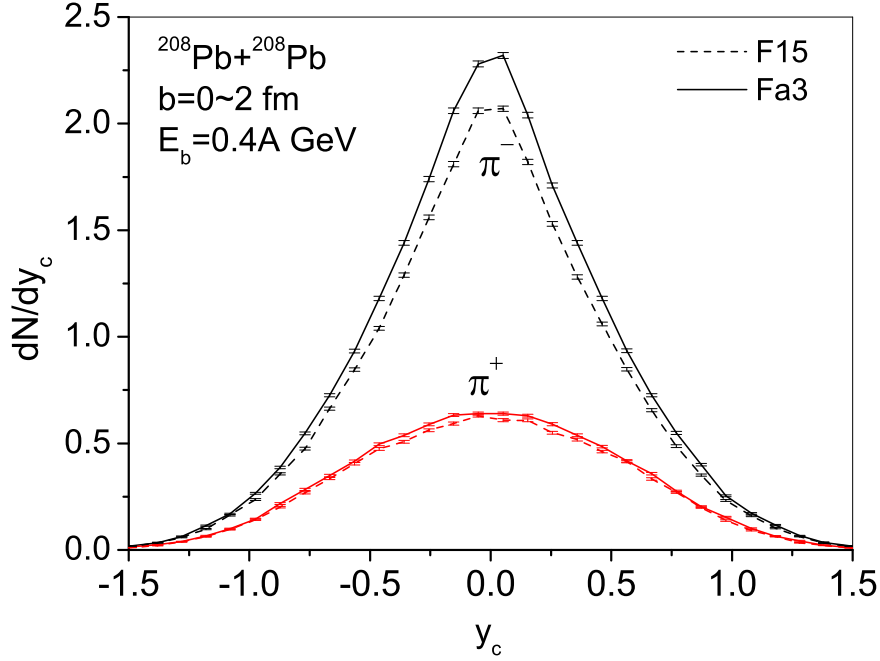}
\caption{The rapidity distributions of $\pi^-$ and $\pi^+$ for
central $^{208}Pb+^{208}Pb$ collisions with different symmetry
potentials (see text). } \label{fig2}
\end{figure}

\begin{figure}
\includegraphics[angle=0,width=0.8\textwidth]{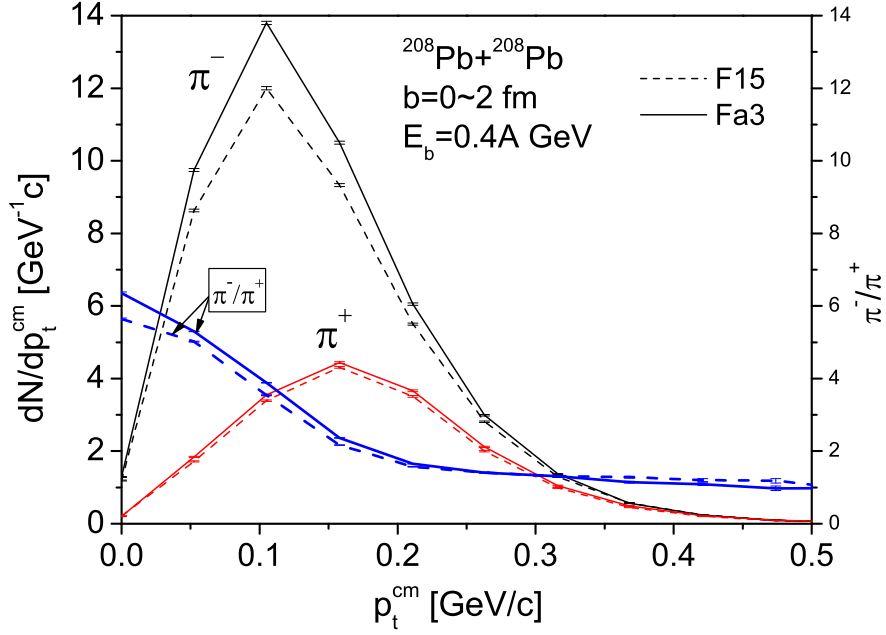}
\caption{Transverse momentum distributions of $\pi^-$ and $\pi^+$
from central $^{208}Pb+^{208}Pb$ collisions at $E_b=0.4A\,{\rm
GeV}$ for different symmetry potentials. The $\pi^-/\pi^+$ ratios
are also shown as a function of transverse momentum.} \label{fig3}
\end{figure}

Figs.\
\ref{fig2} and \ref{fig3} show the rapidity and the transverse
momentum distributions of $\pi^-$ and $\pi^+$ for central ($b=0-2$
fm) $^{208}Pb+^{208}Pb$ collisions at $0.4A\,{\rm GeV}$ as
calculated with the symmetry potentials $F15$ and $Fa3$. Fig.\
\ref{fig3} also shows the corresponding $\pi^-/\pi^+$ ratios. While the
$\pi^-$ yields are quite sensitive to the symmetry potential at
mid-rapidity and at low $p^{\rm cm}_{t}$, the $\pi^+$ do not
exhibit this sensitivity due to the neutron-rich environment \cite{LiB05}. The calculations for the
$^{132}Sn+^{124}Sn$ and $^{96}Zr+^{96}Zr$ reactions are similar to
those in Figs.\ \ref{fig2} and \ref{fig3}. To analyze the
differences in the $\pi^-$ and $\pi^+$ yields and the
$\pi^-/\pi^+$ ratios (as calculated with the four different forms
of the symmetry potential) more quantitatively, we define a
variable $D_{ij}=(X_i-X_j)/X_j$, where the subscripts "$i$" and
"$j$" denote different forms of the symmetry potential. "$X$"
represents the calculated quantity, such as the pion yield or the
$\pi^-/\pi^+$ ratio. Here, $i$ denotes the symmetry potential that
yields a larger $X$ value than the one with the symmetry potential
$j$, i.e., we always choose $X_i > X_j$. Table \ref{tab1} (for
$^{208}Pb+^{208}Pb$, titled as "$Pb208$") and Table \ref{tab2} (
for $^{132}Sn+^{124}Sn$, titled as "$Sn132$") give the results of
$D_{ij}$ for different conditions and physical cuts: the subscript
"tot" indicates the results without any kinematical cut, "$y\sim 0$"
denotes particles emitted at mid-rapidity ($y_c$ in the
center-of-mass system), and the subscript "$p_t\sim 0.1$" means
the particles with transverse momenta $p_t^{\rm cm}\simeq
0.1A\,{\rm GeV}/c$ . Here, the index 1 represents the
"DDH3$\rho\delta^*$" symmetry potential, 2 represents
"DDH$\rho^*$", 3 represents "F15", and 4 is the "Fa3" case.
$D_{max}$ provides the maximum value of all $D_{ij}$. The differences are maximal for
the $D_{23}$ values, these are the differences between the F15 and
DDH$\rho^*$ symmetry potentials, which also exhibit the largest deviations in the density dependence of the symmetry potential at densities $u \sim 1-2.6$ (see Fig. \ref{fig1}). Tables \ref{tab1} and \ref{tab2} demonstrate that the values of $D_{ij}$ for
$\pi^{+}$ are always smaller than those for $\pi^{-}$, this is consistent with Figs.\ \ref{fig2} and \ref{fig3}. By comparison,
we find that the $\pi^-/\pi^+$ ratio at $p_t^{\rm cm}\simeq
0.1\,{\rm GeV}/c$ is most suitable to probe the
symmetry potential because of the larger $\pi^-$ and $\pi^+$
yields in this transverse momentum region. We divide all $D_{ij}$
into two groups in Tables \ref{tab1} and \ref{tab2}. In general, the
values in the first group of $D_{21}$, $D_{43}$ and
$D_{max}$($=D_{23}$) are much larger than those in the second
group of $D_{13}$, $D_{41}$, $D_{24}$. For the former case the
values are generally larger or close to $10$ except for the
$\pi^{+}$, while for the latter one they are all smaller than $10$,
except $D_{13}$ for the $\pi^-/\pi^+$ ratio at $p_t^{\rm cm}\simeq
0.1\,{\rm GeV}/c$ with a value of $10$. Surprisingly, $D_{41}$ lies
in the second group, although Fa3 corresponds to a soft symmetry
potential and DDH3$\rho\delta^*$ corresponds to a stiff symmetry
potential. The reason can be found in Fig.\ \ref{fig1}: the
difference between these two symmetry potentials is not large for
reduced densities $u=1\sim 2$ where most of the pions are produced
\cite{LiQ052}. It means that more accurate measurements are needed
in order to probe the density dependence of the symmetry potential
with pions, except for the extreme cases. Furthermore, from Tables
\ref{tab1} and \ref{tab2} we observe $X_{DDH\rho^*}
> X_{Fa3} > X_{DDH3\rho\delta^*}
>X_{F15}$. This order agrees with the order
of the four forms of the symmetry potentials at baryon densities
$u \simeq 1-2$ (see Fig.\ \ref{fig1}). Baryon densities
of $u \simeq 1-2$ are the most relevant ones for pion
production as we have shown in \cite{Li05,LiQ052}. Comparing the
values $D_{ij}$ shown in Tables \ref{tab1} and \ref{tab2}, we see
that the values $D_{ij}$ are a little smaller for
$^{208}Pb+^{208}Pb$ as compared to $^{132}Sn+^{124}Sn$ with
radioactive beams.

\begin{table}

\caption{Values $D_{ij}$ of the total $\pi^{\pm}$ yields (at
mid-rapidity or $p_t^{\rm cm} \simeq 0.1\,{\rm GeV}/c$) and of the
corresponding $\pi^-/\pi^+$ ratios for various combinations of the
density-dependent symmetry potentials for $^{208}Pb+^{208}Pb$
reactions (see text).}

\begin{tabular}{|l|ccc|ccc|ccc|}
\hline\hline
"Pb208"  & $\pi^{-}_{tot}$ & $\pi^{-}_{y\sim0}$ & $\pi^{-}_{p_t\sim 0.1}$ & $\pi^{+}_{tot}$ & $\pi^{+}_{y\sim 0}$ & $\pi^{+}_{p_t\sim 0.1}$ & ($\frac{\pi^-}{\pi^{+}})_{tot}$ & ($\frac{\pi^-}{\pi^{+}})_{y\sim 0}$ & ($\frac{\pi^-}{\pi^{+}})_{p_t\sim 0.1}$ \\
\hline
$D_{21}$ [\%] & $14$ & $13$&$16$ & $9$ & $11$ & $9$ & $5$ & $2$ & $6$\\
$D_{43}$ [\%] & $12$ & $11$&$15$ & $5$ & $3$ & $5$ & $7$ & $8$ & $10$\\
$D_{max}$ [\%] & $20$ & $18$ &$24$ & $10$ & $12$ & $9$ & $8$ & $5$ & $18$\\
\hline\hline
$D_{13}$ [\%] & $5$ & $5$&$7$ & $1$ & $1$ & $0$ & $4$ & $3$ & $10$\\
$D_{41}$ [\%] & $6$ & $6$&$8$ & $4$ & $2$ & $5$ & $3$ & $4$ & $0$\\
$D_{24}$ [\%] & $7$ & $6$ &$8$ & $5$ & $8$ & $4$ & $2$ & $5$ & $7$\\
\hline\hline

\end{tabular}
\label{tab1}
\end{table}

\begin{table}

\caption{The same as in Table \ref{tab1} but for
$^{132}Sn+^{124}Sn$ reactions.}

\begin{tabular}{|l|ccc|ccc|ccc|}
\hline\hline
"Sn132"  & $\pi^{-}_{tot}$ & $\pi^{-}_{y\sim 0}$ & $\pi^{-}_{p_t\sim 0.1}$ & $\pi^{+}_{tot}$ & $\pi^{+}_{y\sim 0}$ & $\pi^{+}_{p_t\sim 0.1}$ & ($\frac{\pi^-}{\pi^{+}})_{tot}$ & ($\frac{\pi^-}{\pi^{+}})_{y\sim 0}$ & ($\frac{\pi^-}{\pi^{+}})_{p_t\sim 0.1}$ \\
\hline
$D_{21}$ [\%] & $12$ & $11$&$13$ & $5$ & $6$ & $6$ & $6$ & $5$ & $7$\\
$D_{43}$ [\%] & $11$ & $10$&$13$ & $4$ & $4$ & $3$ & $7$ & $6$ & $10$\\
$D_{max}$ [\%] & $19$ & $17$ &$21$ & $7$ & $8$ & $6$ & $11$ & $9$ & $18$\\
\hline\hline
$D_{13}$ [\%] & $6$ & $5$ &$7$ & $1$ & $2$ & $0$ & $5$ & $3$ & $10$\\
$D_{41}$ [\%] & $5$ & $5$ &$6$ & $3$ & $2$ & $4$ & $2$ & $3$ & $0$\\
$D_{24}$ [\%] & $6$ & $6$ &$7$ & $2$ & $3$ & $2$ & $4$ & $2$ & $8$\\
\hline\hline

\end{tabular}
\label{tab2}
\end{table}

Probes being sensitive to the density dependence of the symmetry
potential at subnormal densities have been studied extensively in
recent years. Usually a form $(\rho/\rho_{0})^{\gamma}$ is assumed
for the density dependence of the symmetry potential \cite{Tsa04}. From Fig.\
\ref{fig1} it follows that the behavior of the symmetry potential
at subnormal and high densities are quite different. To gain
information on the symmetry potential in both, the high and the
subnormal density regions simultaneously, observables from the
same reaction system are needed, which are sensitive to both density regions. Most of the free nucleons, intermediate mass
fragments and light charged particles are emitted in the expansion
stage, that is, at subnormal densities. Therefore, we further
study the ratios of free neutrons and protons, the isotope
distributions of light and intermediate mass fragments, as well as the ratio of
emitted tritons and $^{3}$He which is also proposed as a
sensitive probe for the symmetry potential at subnormal densities
\cite{Zhang05}.  A conventional phase-space coalescence model \cite{Kru85} is
used to construct the clusters, in which the nucleons with
relative momenta smaller than $P_0$ and relative distances smaller
than $R_0$ are considered to belong to one cluster. In this work
$P_0$ and $R_0$ are chosen to be $0.3\,{\rm GeV}/c$ and $3.5$ fm,
respectively. Another set, $P_0=0.25\,{\rm GeV}/c$ and $R_0=3.0$
fm, alters the yields of free nucleons and clusters strongly, but
it does not change the dependence of the $t/^{3}$He ratios on the
symmetry potentials. The freeze-out time is taken to be $150\ {\rm
fm}/c$. In Ref. \cite{Zhang05} a modified coalescence model
\cite{Neu03} was adopted. It was found that a modification
of the coalescence criteria changes the yields of light charged
particles but it does not change the $t/^{3}$He ratio.

\begin{figure}
\includegraphics[angle=0,width=0.8\textwidth]{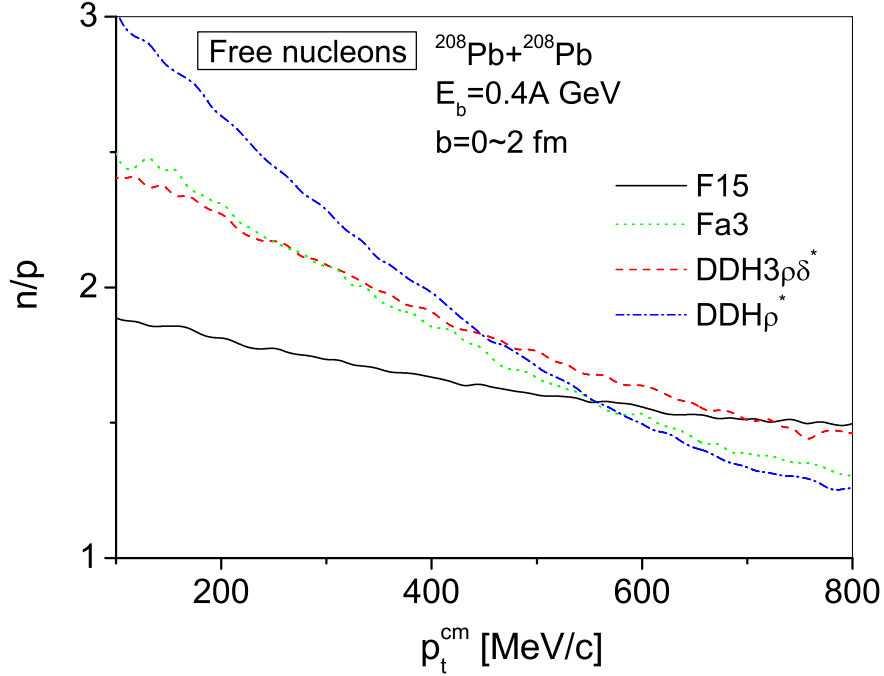}
\caption{Transverse momentum distributions of free neutron/proton
ratios for different density-dependent symmetry potentials.}
\label{fig4}
\end{figure}

Fig.\ \ref{fig4} shows the transverse momentum distribution of the
$n/p$ ratios of emitted unbound nucleons as calculated with the
four different forms of the symmetry potential. The $n/p$ ratio
obviously depends strongly on the choice of the symmetry
potential. In the low transverse momentum region the $n/p$ ratio
is the largest for DDH$\rho^*$ and the smallest for F15. Obviously,
nucleons with low transverse momenta are mainly emitted from the
low density region. From Fig.\ \ref{fig1} it follows that the
difference between Fa3 and DDH3$\rho\delta^*$ is small at
densities $\rho<\rho_0$. Correspondingly, the $n/p$ ratios
calculated with Fa3 and DDH3$\rho\delta^*$ are also very close at
lower transverse momenta. For emitted nucleons with transverse
momenta larger than $\sim 700\ {\rm MeV}/c$, the $n/p$ ratios in
the Fa3 and DDH3$\rho\delta^*$ cases are close to those in the
DDH$\rho^*$ and F15 cases, respectively. This reflects the fact that
$E_{\rm sym}^{\rm pot}$ for Fa3 (DDH3$\rho\delta^*$) is close to
DDH$\rho^*$ (F15) when $\rho
>\rho_{0}$. Free nucleons with transverse momenta larger than $\sim\ {\rm 700} MeV/c$ are mainly squeezed out
from higher densities \cite{Bass94}. Thus, the different behavior of the $n/p$
ratio as a function of {\it transverse momentum} goes in line with
the behavior of the symmetry potential at low and high {\it
densities} as shown in Fig.\ \ref{fig1}. The relative difference
between the $n/p$ ratios for transverse momenta larger than $\sim
700\ {\rm MeV}/c$ as calculated with the soft and stiff symmetry
potentials is on the same order as for the $\pi^{-}/\pi^{+}$
ratios.

\begin{figure}
\includegraphics[angle=0,width=0.8\textwidth]{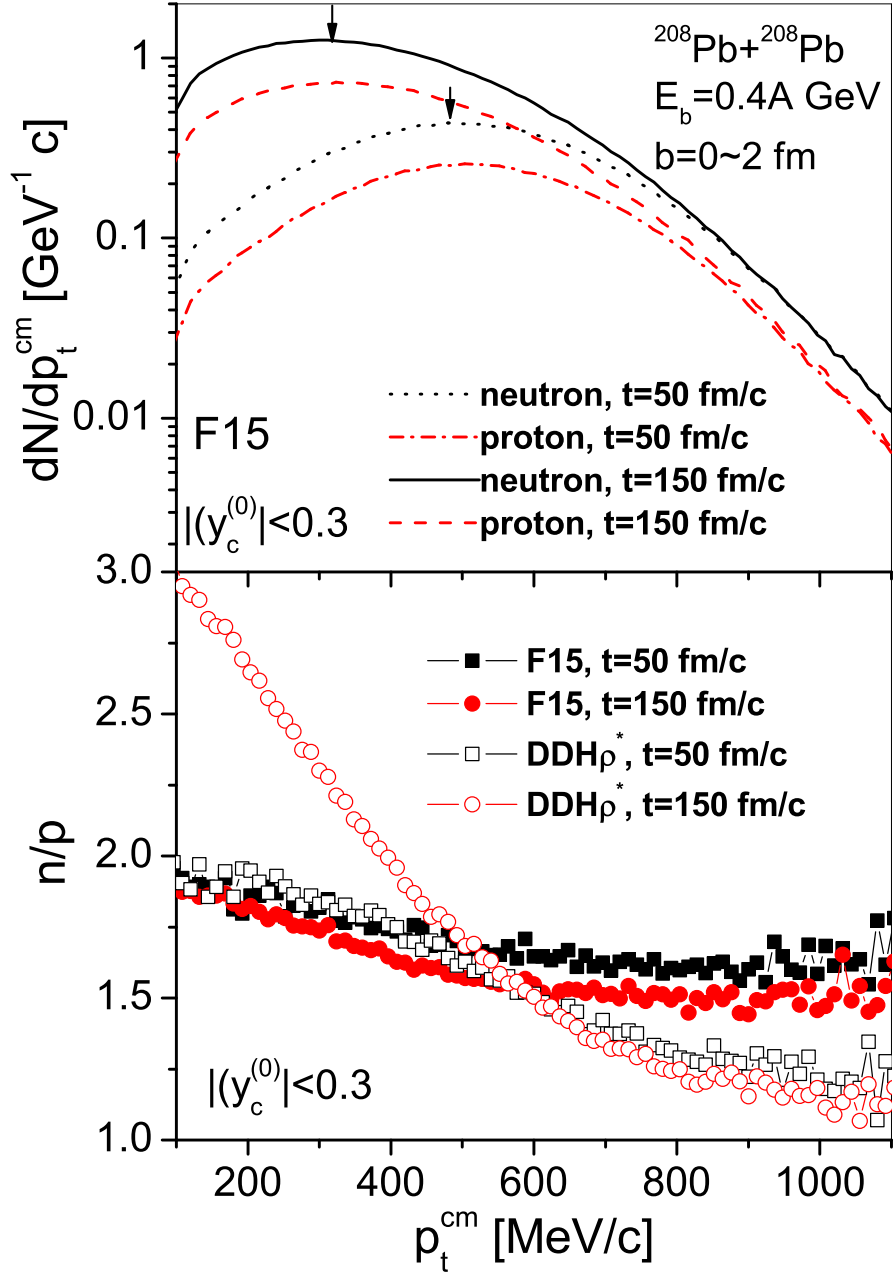}
\caption{Transverse momentum distributions of free neutrons and
protons (upper plot) as well as the corresponding $n/p$ ratios
(lower plot) at mid-rapidity ($|(y_c^{(0)}|<0.3$) and at reaction
times $t=50$ and $150\ {\rm fm}/c$. The F15 and DDH$\rho^*$
symmetry potentials are chosen. The arrows in the upper plot show
the maxima of the distributions.} \label{fig5}
\end{figure}

To see the different emission times for nucleons with high and low
transverse momenta, Fig.\ \ref{fig5} shows the transverse momentum
distributions of free neutrons and protons (upper plot) as well as
the corresponding $n/p$ ratios (lower plot), emitted from the
mid-rapidity region ($|(y_c^{(0)}|=|y_c/y_{beam}|<0.3$) at
reaction times $t=50$ and $150\ {\rm fm}/c$. In the upper plot of
Fig.\ \ref{fig5} only the results for F15 are shown while the ratios for both the F15 and DDH$\rho^*$ symmetry
potentials are compared in the lower plot. Nucleons with large transverse momenta
are emitted in the early stage of the reaction and nucleons with
small transverse momenta are emitted at later time. Furthermore, we see
an obvious change of the transverse momentum distribution of the
$n/p$ ratios with time, especially in the low transverse momentum
region, in particular in the case of the soft symmetry potential DDH$\rho^*$.
This can easily be understood from Fig.\ \ref{fig1}: in the cases
of the soft symmetry potentials $DDH\rho^*$ and Fa3, neutrons are
strongly attracted into the high density region by the symmetry
potential at the compression stage. Within the expansion period,
neutrons are driven to lower densities, eventually leading to the
subnormal densities. At subnormal densities the soft symmetry potential
energy is higher than the stiff one. Hence, more neutrons with
lower transverse momenta are emitted for the cases with soft
symmetry potentials $DDH\rho^*$ and Fa3 at the later stage (at
$150\ {\rm fm}/c$) as compared to the cases with the stiff
symmetry potentials. For the stiff symmetry potential case, there
is no significant change of the transverse momentum distributions of the
$n/p$ ratio from $t=50\ {\rm fm}/c$ to $150\ {\rm fm}/c$. In the
high transverse momentum region, the $n/p$ ratios for both cases
decrease with time during the expansion phase, which happens
mainly due to the extra Coulomb effect on the protons.

\begin{figure}
\includegraphics[angle=0,width=0.8\textwidth]{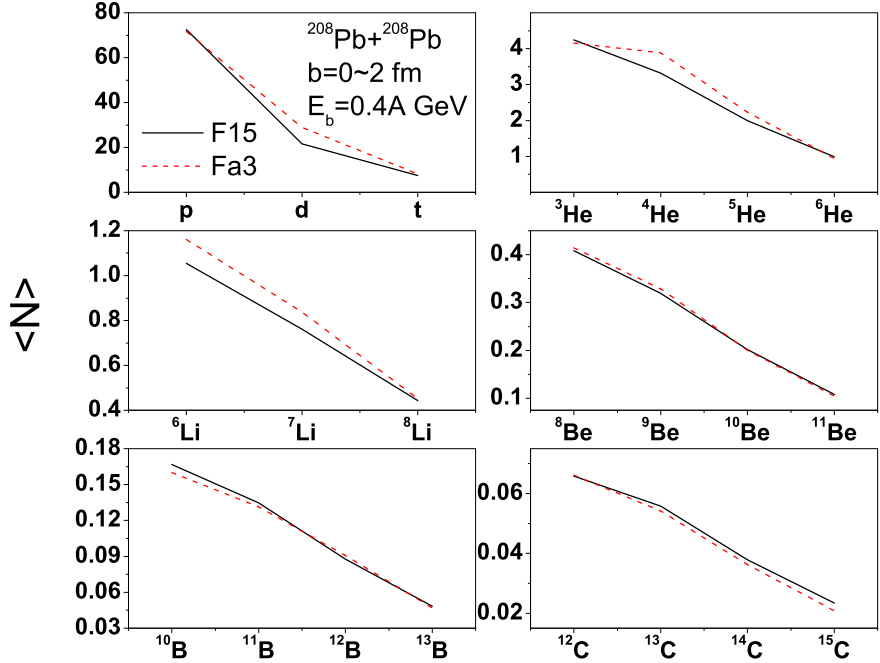}
\caption{Isotope distributions of H, He, Li, Be, B and C with the
symmetry potentials F15 and Fa3 (see text). } \label{fig6}
\end{figure}

However, experimentally it is difficult to measure the spectra of
neutrons with high precision. Therefore, the production of light charged particles
with large but different N/Z ratios as well as isotope distributions
for some intermediate mass fragments could be more useful
\cite{LiQ01,Chen0304}.  Fig.\ \ref{fig6} shows the isotope
distributions of $^{1-3}$H, $^{3-6}$He, $^{6-8}$Li, $^{8-11}$Be,
$^{10-13}$B and $^{12-15}$C in central Pb+Pb collisions as
calculated with the symmetry potentials F15 and Fa3, respectively. In general,
the differences between the isotope yields for the F15 and Fa3
cases are smaller than the differences in the $n/p$ ratio of
emitted nucleons. However, as
the isotopes become more neutron-rich, the corresponding yields
decrease fast. Secondly, with an increase of the neutron number in
the light isotopes of H, He and Li, the differences between the
yields calculated with F15 and Fa3 enlarge (diminish) for
symmetric (asymmetric) isotopes, while for the heavier Be, B, and
C elements the isotope distributions are almost the same for F15
and Fa3. This can be understood based on the density dependence of
the symmetry potentials shown in Fig.\ \ref{fig1}: for He and Li
isotopes, a weak enhancement of isospin symmetric isotopes
with Fa3 is shown but this is not observed for neutron rich
isotopes. This is because the $n/p$ ratio for free nucleons, as
calculated with Fa3, is much larger than that with F15, which leads
to the reduction of the N/Z ratios for H, He, and Li when
calculated with Fa3.

\begin{figure}
\includegraphics[angle=0,width=0.8\textwidth]{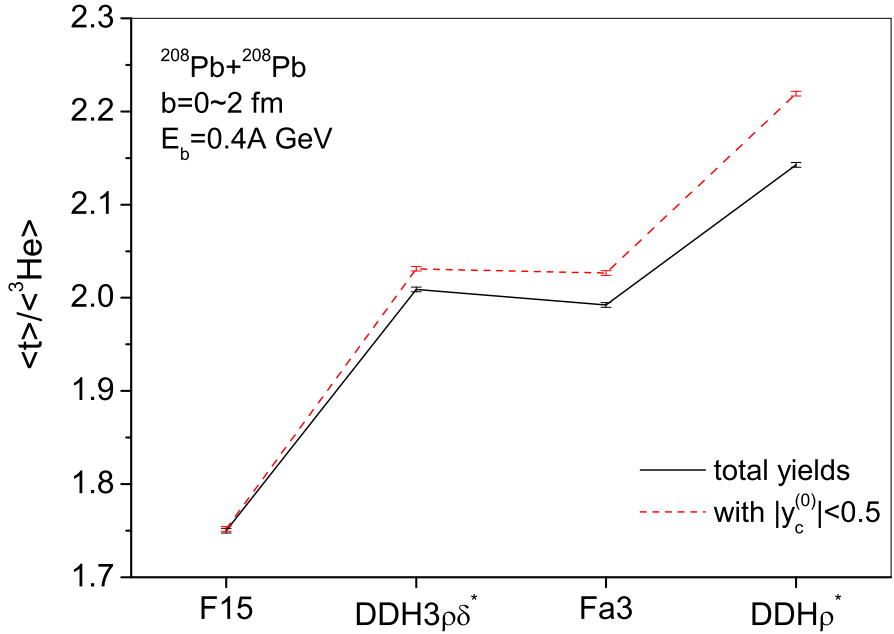}
\caption{The ratios of the triton and $^3$He yields with and
without rapidity cut as calculated with four different forms of
symmetry potentials.} \label{fig7}
\end{figure}

The light charged particles like triton and $^3$He ought to be
particularly noted, because they exhibit a considerable difference in the
isospin asymmetry, as well as large yields in heavy ion collisions at SIS energy, $E_b\lesssim 1A\ {\rm GeV}$. Both the FOPI \cite{Pog95} and the EOS \cite{Lis95}
collaborations have found that - in central Au+Au collisions - $^3$He clusters
have a systematically larger mean kinetic energy than tritons up
to bombarding energies of $400A$ MeV. This trend is also found in
our calculations. Fig.\ \ref{fig6} shows the yield
of tritons is larger than that of $^3$He. To see the symmetry
potential effect more clearly, Fig.\ \ref{fig7} shows the
$t/^{3}$He ratios as calculated with the four different forms of
the symmetry potential used in this work. The dashed line is for
the case with the rapidity cut and the solid line is for the case
without any kinematical cut. Fig.\ \ref{fig7} shows that the $t/^{3}$He
ratios are more strongly enhanced with a softer symmetry potential
at mid-rapidity $|y_c^{(0)}|<0.5$ than without any
cut. This is because neutron-richer matter is formed in
the mid-rapidity region with an increasingly softer symmetry
potential. Thus, more (less) tritons ($^3$He) are produced with a
softer symmetry potential as compared to the case without any cut.
The magnitudes of the calculated $t/^3$He ratios can be ordered as follows:
$X_{DDH\rho^*} > X_{DDH3\rho\delta^*} > X_{Fa3} > X_{F15}$. This
is a little different from the ordering of the $\pi^{-}/\pi^{+}$ ratios. However,
the difference of $t/^{3}$He is very small between Fa3 and
DDH3$\rho\delta^*$. $D_{max}$ is about $22\%$ for
the $t/^3$He ratio in full phase space ($4\pi$) and $D_{max}
\simeq 27\%$ for the $|(y_c^{(0)}|<0.5$ case. This is larger than
in the case of pions as given in Tables \ref{tab1} and \ref{tab2}.
Comparing with Fig.\ \ref{fig1} it becomes clear that the ordering of
the $t/^3$He ratios for the four different forms of the symmetry
potential corresponds to the ordering of the $E_{\rm
sym}^{pot}$ at subnormal densities! This holds, regardless of the
behavior at $u>1$. Therefore, we consider the $t/^{3}$He ratio
as a sensitive probe of the density dependence of the symmetry
potential at subnormal densities.

In summary, we have
investigated the influence of different density dependences of the
symmetry potential on charged pion yields, the $\pi^{-}/\pi^{+}$
ratio, the $n/p$ ratio for free nucleons, the isotope distribution
of H, He, Li, Be, B, C and the $t/^3$He ratio for
$^{208}Pb+^{208}Pb$ ($\delta \simeq 0.212$), $^{132}Sn+^{124}Sn$
($\delta \simeq 0.218$), and $^{96}Zr+^{96}Zr$ ($\delta \simeq
0.167$) at $E_b=0.4A\,{\rm GeV}$, respectively. We have shown that
the negatively charged pion yields, and the $\pi^-/\pi^+$ ratios
at low momenta are very sensitive to the density dependence of the
symmetry potential at high densities, while the $n/p$ ratio, i.e., for
free nucleons, and the $t/^{3}$He ratio are sensitive to the potential at
subnormal densities. The $n/p$ ratios at high transverse momenta show sensitivity
to the behavior of the symmetry potential at high densities. To obtain the full information on the density dependence of
the symmetry potential, several sensitive probes
at both high and low densities were analyzed simultaneously. The comparison
of the results of different systems shows that it is
suitable to use stable but highly isospin-asymmetric systems
to study the effects of different symmetry
potentials.

Here, it should be mentioned again that concrete values for the yields
of emitted pions, neutrons, protons, tritons, and $^{3}$He may be
influenced by the uncertainty of the isospin-independent (and the
momentum dependent) part of the EoS. However, we expect that the
dependence of the $n/p$ , $\pi^{-}/\pi^{+}$, and $t/^{3}$He ratios
on the symmetry potential is not strongly affected.

\section*{Acknowledgments}
Q. Li thanks the Alexander von Humboldt-Stiftung for a fellowship.
This work is partly supported by the National Natural Science
Foundation of China under Grant No.\ 10255030, by GSI, BMBF, DFG, and VolkswagenStiftung.

\end{document}